# MASTER optical detection of the first LIGO/Virgo neutron stars merging GW170817.

V.M. Lipunov[1,2], E. Gorbovskoy[2], V.G. Kornilov[1,2], N. Tyurina[2], P. Balanutsa[2], A. Kuznetsov[2], D. Vlasenko[1,2], D. Kuvshinov[1,2], I. Gorbunov[2], D.A.H. Buckley[3], A.V. Krylov[2], , R. Podesta[8], C. Lopez[8], F. Podesta[8], H. Levato[9], C. Saffe[9] , C. Mallamachi[10], S. Potter[3], N.M. Budnev[5], O. Gress[5,2], Yu. Ishmuhametova[5], V. Vladimirov[2], D. Zimnukhov[2], V. Yurkov[7], Yu. Sergienko[7], A. Gabovich[7] , R. Rebolo[4], M. Serra-Ricart[4], G. Israelyan[4], V. Chazov[2], Xiaofeng Wang[11], A. Tlatov[6], M.I. Panchenko[2]

[1]*M.V. Lomonosov Moscow State University, Physics Department, Leninskie gory, GSP-1, Moscow, 119991, Russia*
[2]*M.V. Lomonosov Moscow State University, Sternberg Astronomical Institute, Universitetsky pr., 13, Moscow, 119234, Russia*
[3]*South African Astrophysical Observatory, PO Box 9, 7935 Observatory, Cape Town, South Africa*
[4]*Instituto de Astrofacuteisica de Canarias Via Lactea, s/n E38205 - La Laguna (Tenerife), Spain*
[5]*Irkutsk State University, Applied Physics Institute, 20, Gagarin blvd, 664003, Irkutsk, Russia*
[6]*Kislovodsk Solar Station of the Main (Pulkovo) Observatory RAS, P.O.Box 45, ul. Gagarina 100, Kislovodsk 357700, Russia*
[7]*Blagoveschensk State Pedagogical University, Lenin str., 104, Amur Region, Blagoveschensk 675000,*
[8]*Observatorio Astronomico Felix Aguilar (OAFA) , National University of San Juan, Argentina*
[9]*Instituto de Ciencias Astronomicas, de la Tierra y del Espacio (ICATE), San Juan, Argentina*
[10]*National University of San Juan, Argentina*
[11]*Tsinghua University, China*

\* E-mail: lipunov2007@gmail.com (VML)

## ABSTRACT

Following the reported discovery of the gravitational-wave pulse GW170817/ G298048 by three LIGO/Virgo antennae (Abbott et al., 2017a), the MASTER Global Robotic Net telescopes obtained the first image of the NGC 4993 galaxy after the NS+NS merging. The optical transient MASTER OTJ130948.10-232253.3/SSS17a was later found, which appears to be a kilonova resulting from a merger of two neutron stars. In this paper we report the independent detection and photometry of the kilonova made in white light and in B, V, and R filters. We note that luminosity of the discovered kilonova NGC 4993 is very close to another possible kilonova proposed early GRB 130603 and GRB 080503.

1.INTRODUCTION

There have been several reasons to expect that neutron-star mergers must be accompanied by electromagnetic radiation before, during, and after the gravitational-wave pulse. Blinnikov et al. (1984) were the first to associate the gamma-ray bursts with the result of the explosion of the neutron star during a merger.

Lipunov & Panchenko (1996) showed that a merger of two magnetized neutron stars (or one such star if the second component is a black hole) can be expected to be accompanied by a non-thermal electromagnetic burst, the precursor of a pulsar. Later, Hansen et al. (2001) illustrated the idea of Lipunov & Panchenko (1996) using a detailed electromagnetic model.

After a merger (Clark et al. 1979 ) part of the matter from the neutron star can be ejected and this may result in radioactive decay of the synthesized heavy elements – the so-called kilonova (Li & Paczynski 1998; Metzger et al. 2010 ;Tanvir et al. 2013 ; Berger et al. 2013). On the other hand, a rapidly rotating self-gravitating object - a spinar - may be a source of strong, long burst of electromagnetic radiation (Lipunova & Lipunov 1998; Lipunov & Gorbovskoy 2008; Lipunova et al. 2009 ). The spinar may represent an intermediate phase and its rotational evolution may end up with the formation of a highly magnetized heavy neutron star, namely a magnetar.

The MASTER Global Robotic Net has been taking an active part in the follow-up searches for optical afterglows of all detected LVC events (Lipunov et al. 2017a,b) since the detection of the first gravitational-wave event with the advanced LIGO interferometers (Abbot et al., 2016a,b).

On 17th August 2017, von Kienlin et al (2017) reported a short (2 sec long) gamma-ray burst recorded by the Gamma Burst Monitor mounted on the Fermi satellite (Fermi GBM trigger 524666471) at 12:41:06.47 UT, which occurred two seconds after the detection of the gravitational-wave event (Essick et al. 2017, Connaughton et al. 2017a ). Later, Savchenko et al. (2017) also found a short and relatively weak transient with a S/N >3, coincident with GBM trigger (Abbott et al.2017b).

On the next day the 1-m SWOPE telescope at Las Campanas Observatory was the first to report the new optical source, SSS17a, located 5.3 arcsec E and 8.8 arcsec N of an S0 galaxy in the NGC 4993 / ESO 508-G018 group (Coulter et al., 2017), at a distance of ~40 Mpc (Cook et al. 2017, Kasliwal et al. 2017) also independently discovered by MASTER auto-detection system (Lipunov et al. 2017c), and confirmed by other telescopes in different filters in first hours (Abbott et al.2017a,b). EM partners started its broad-band investigations (Chambers et al.2017,Drout et al 2017, Shara et al 2017, Abbott 2017a,b).

MASTER Global Robotic Net observations of the G298040/GW170817 error region started with the MASTER-SAAO, in South Africa, at 2017-08-17 17:06:47UT (0.4 day after trigger), then continued in MASTER-IAC, in Spain, at 2017-08-17 20:29:26UT, and in Argentina with the MASTER-OAFA, which obtained the first image of NGC4993 at 22:54:18UT using the very wide field cameras, but did not detect any optical transient, to a limit of with m=15.2 (5 sigma in white light and corrected for Galaxy extinction).

Another words, MASTER-OAFA obtained the first image of the NGC 4993 galaxy after the NS+NS merging.

The optical transient MASTER OTJ130948.10-232253.3/SSS17a was independently discovered by MASTER-OAFA auto-detection system during inspection by main MASTER-OAFA telescope at 23:59:54UT (Lipunov et al.2017c,d,e; Abbott et al. 2017a,b).

These discoveries and subsequent observations showed quite conclusively that on 17 August 2017 astronomers observed a merger of two neutron stars in the galaxy NGC 4993 and its afterglow and this was the first time that such event was observed, not only in gravitational-waves, but also over the electromagnetic spectrum ranging from gamma rays, X-rays, ultraviolet, optical and infrared radiation. This first detection of an electromagnetic counterpart comes only 2 years after the first confirmed detection of a gravitational wave event. This is compared to the three decades in took for the equivalent detection of gamma ray bursts at other wavelengths.

## 2. MASTER-NET IN LIGO/VIRGO FOLLOW-UP

The MASTER Global Robotic Net (Lipunov et al. 2010) consists of 2 classes of instruments: the main MASTER system of twin 40-cm wide field (2x4 square degrees, 1 pixel = 1.9 arcsec) optical robotic telescopes and two very wide field cameras (MASTER-VWF). MASTER-VWF is a very fast camera capable of obtaining up to three images per second and equipped with an 82 mm aperture F/1.2 lens giving a 16 x 24 deg = 384 sq. degree field of view, capable of detecting objects down to a limiting magnitude of $12^m$ per single 5-s exposure image. Each of the MASTER-Net observatories are equipped with two MASTER-VWF cameras, which are fixed on with the same mount as the main MASTER-II telescopes and together cover a 32x24 degree area centered on the main MASTER telescope direction. For more details see Lipunov et al. (2010), Kornilov et al. (2012), Gorbovskoy et al. (2010, 2013) and the above paper.

MASTER nodes are located at the following observatories:
MASTER-Amur, MASTER-Tunka, MASTER-Kislovodsk, MASTER-Tavrida, MASTER-Ural (Russia), MASTER-SAAO (South Africa), and MASTER-IAC (Spain, Canarias, Teyde observatory), MASTER-OAFA (San Juan National University observatory in Argentina) (see Figure1).

All MASTER observatories are equipped with identical own photometers (two full frame CCD cameras, B, V,R, I Johnson-Bessel filters, and two linear polarizers), and are controlled by identical own software.The use of identical equipment allows us to have up to 24 hours continuous observations of optical counterparts of transients in identical photometric systems. Hence combining photometric data for different transients observed from different parts of the MASTER Net is a proven astronomical process (see the last result in Troja et al. 2017a, Lipunov et al., 2016).

The large field of view of every MASTER main telescopes allows us to use a large number (1,000 - 5,000) reference stars for photometric reductions. As a result, the photometric errors can be minimized, using the large catalogs likeTycho II and USNO-B1, .

We have had succesfull experience in the aLIGO follow up campaign during GW150914 investigation (Abbott et al.2016a,b, Lipunov et al. 2017a,b). MASTER made the most input to the optical support of this event (see Table 1 and Section 5 in Abbott et al. 2016c).

### 3. GW170817 MASTER OBSERVATIONS.

The first alert of the LIGO/Virgo G298048 event arrived when it was daytime at most of the MASTER-Net observatories. Only at MASTER-Amur, in the Russian Far East, there was a night time, but observations were prevented by unfavorable weather conditions .

MASTER Global Robotic Net started observing the LIGO/Virgo G298048 error field at 2017-08-17 17:06:47 UT (Lipunov et al. 2017c,d,e).

MASTER-SAAO automatically began to observe part of the common area of the initial LIGO BAYESTAR error region received by the socket connection (Essick et al. 2017) and the Fermi GBM error box (Connaughton et al., 2017) immediately after sunset (Sun altitude <12deg), at 2017-08-17 17:06:47UT = 2017-08-17.71304 (JD = 2457983.21304398), i.e. 15,943 sec after the LVC trigger time (12:41:04 UT), MASTER-IAC began GW170817 initial BAYESTAR map inspection at 2017-08-17 20:29:26UT, see Figure 1.

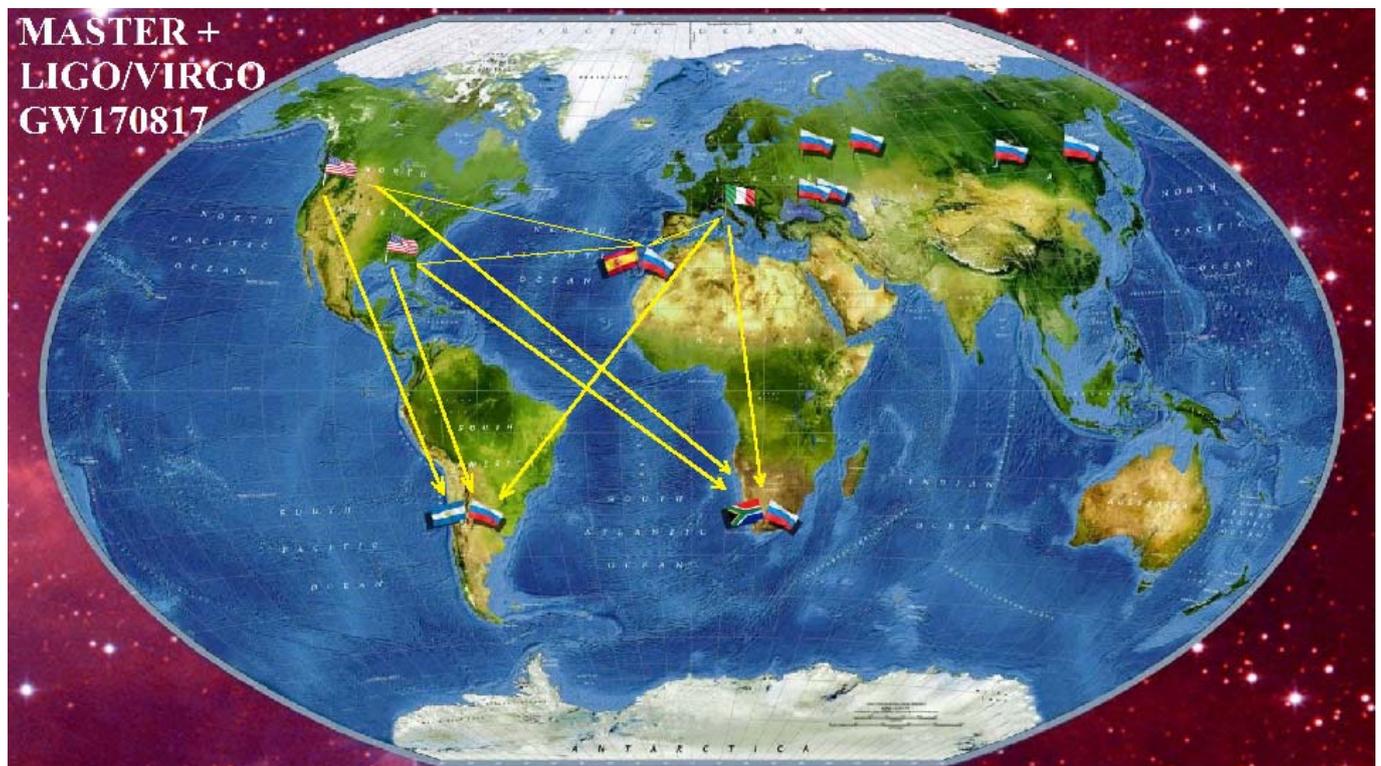

**Figure 1.** MASTER Global Robotic Net and LIGO/VIRGO collaboration interaction in GW170817.

The first co-added unfiltered images (3x180 s images) have a limiting magnitude of $19.8^m$. This first stacked image also covered the most probable IceCube candidate N4 (Bartos et al. 2017 ). However, no optical transients were found in this error region.

The MASTER-SAAO telescopes then continued to observe the initial Fermi and LVC common error area and all Ice Cube candidates. The unfiltered limiting magnitudes and fields can be found in GraceDB and the coverage map is available at Figure 2 and Supplement 1. No optical transients were found during these observations.

The localization map of LIGO/Virgo G298048 (LIGO Scientific Collaboration and Virgo Collaboration (Singer et al.) 2017c) was received at a time when the entire new small error field (it was small compared to the previous one, but still spanned a 125 sq. degree area (3 sigma, i.e. 99.7% region area) was below the horizon for both MASTER-SAAO in South Africa and MASTER-IAC in the Canary Islands. Only for the MASTER-OAFA telescope in Argentina the error field was above the horizon, but it was still daytime at that time, when the message was received.

The MASTER-OAFA, located at Observatorio Astronomico Felix Aguilar (OAFA, National University of San Juan, Argentina) also with two MASTER-VWF cameras, began imaging the new BAYESTAR-HLV (Singer,Price 2016,Singer et al.2016) localization map of LIGO/Virgo G298048 (LIGO Scientific Collaboration and Virgo Collaboration 2017a,b,c) at 2017-08-17 22:54:18 UT, immediately after sunset. Observations started for the first field at RA, DEC = 12h 59m 00.00s -19d 59m 38.00s.

The main MASTER telescope imaged BAYESTAR-HLV localization map that did not, unfortunately, cover NGC 4993 region. But MASTER-OAFA VFW cameras (with larger field of view) obtained the first image of the NGC 4993 galaxy after the NS+NS merging (Lipunov et al., 2017c).

4. MASTER OTJ130948.10-232253.3/ Sss17a IN NGC 4993 OBSERVATIONS.

There are two MASTER Very Wide Field cameras (combined FOV = 760 square degrees, 22 arcsec/pixel) on the same MASTER-OAFA mount. As a result, we have a large series (a "video") of 5-sec MASTER-VWF camera images taken without time gaps and covering the entire LIGO/Virgo *BAYESTAR-HLV* G298048 error box including the NGC 4993 galaxy (Figure 2a).

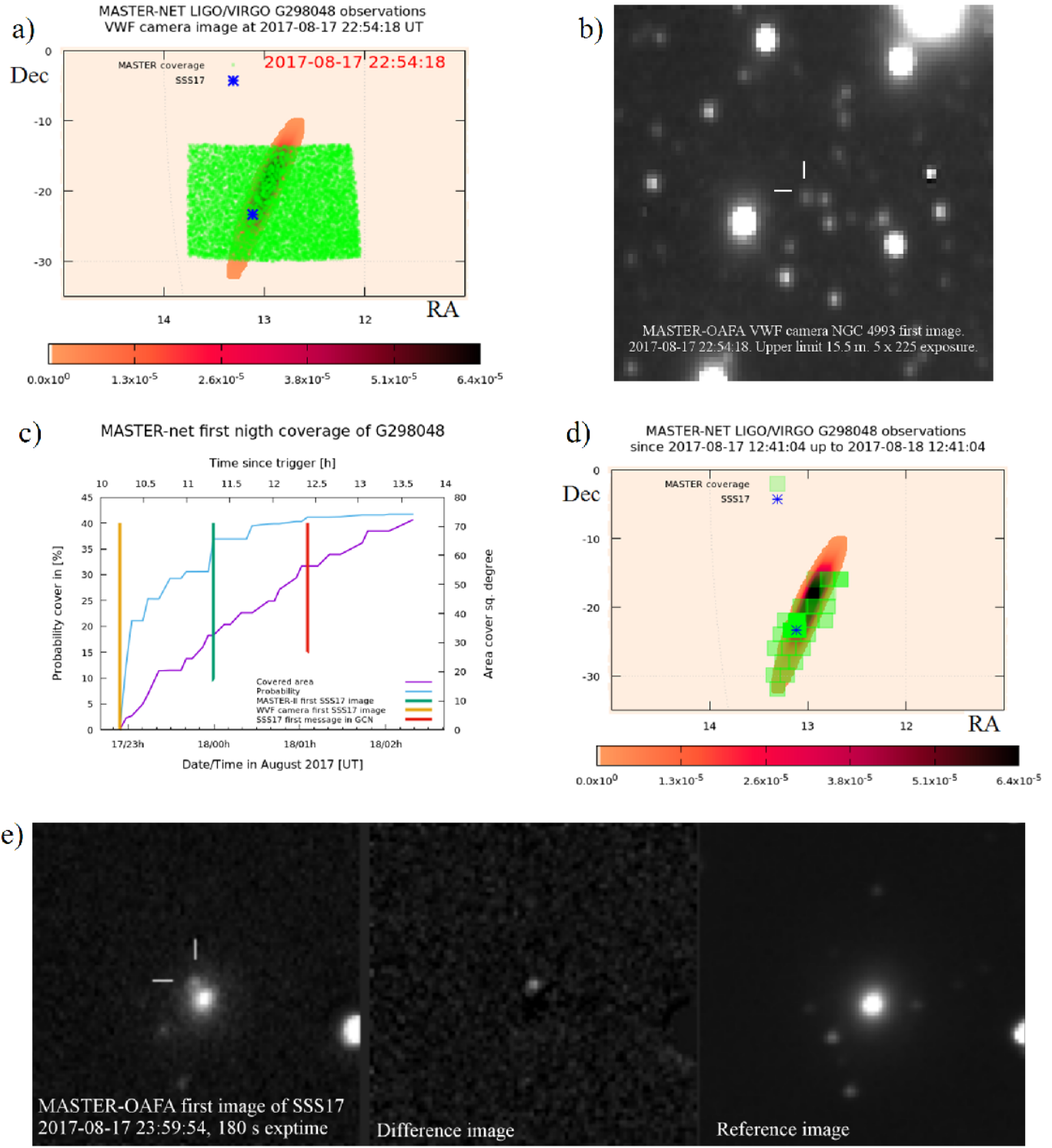

**Figure 2.**

**2a)** MASTER-OAFA Very Wide Field camera coverage of GW170817 error-field starting from 2017-08-17 22:58:48 UT = 2017-08-17 22:54:18UT. The color scale in panels demonstrates G298048 probability distribution. The animation is available at Supplement 1.

**2b)** The first image of the NGC4993 galaxy, 10 hours after the LVC GW170817/G298048 trigger. The image was taken with the MASTER Very Wide Field Camera (VWFC; Lipunov et al., 2010) on the MASTER- OAFA telescope (Argentina).

**2c)** The dynamics of MASTER-Net LIGO/Virgo G298048 BAYESTAR-HLV error area inspections during the first night after the GW170817 (G298048) trigger. The blue line shows the growth of the coverage of the NGC 4993 galaxy where possible KiloNova MASTER OT J130948.10-232253.3/Sss17a was automatically observed by MASTER-VWF cameras and MASTER-OAFA telescope 2.5 and ~1.5 h, respectively, before SWOPE telegram (Coulter et al., 2017)

**2d)** MASTER-Net inspection of LIGO/Virgo G298048 BAYESTAR-HLV error area. The first-night coverage map by MASTER telescope with the limiting magnitude of 20.5.

**2e)** The first MASTER-OAFA image of the kilonova SSS17a in the galaxy NGC 4993, taken starting from 2017-08-17 23:59:54 UT (exposure = 180 sec), 40.73 kilosec after the LVC GW170817/G298048 trigger time. The left panel shows the image obtained after subtracting 75 % of the reference image.

The MASTER-VWF cameras produce a huge data flow of 200 Gb/day, making it impossible for us to store all single image frames for a long time. All single (5 sec) VWF camera images obtained during main MASTER instrument exposure (typically 180 sec) and CCD readout (~ 30 sec) are automatically co-added and archived.

The MASTER-OAFA telescope conducted an inspection survey of the LIGO/Virgo G298048 field quite close (< 10 deg) to the NGC 4993 position and therefore this position is covered by almost all co-added VWF camera images obtained during this survey. In order to obtain the deepest early observation of NGC 4993, we additionally stacked the first six co-added sets of images, taking into account the fact that the region happens to be on different parts of the frames. We thus obtain an extra stacked co-added image of NGC 4993 composing of 225 5-sec single-exposures by MASTER-VWF images with mlim ~ 15.5 mag l (Figure 2b).

The galaxy NGC 4993 can be seen in these co-added images, beginning from 2017-08-17 22:54:18 UT. For our analysis we also used an archival reference image obtained during the previous nights, with the same limiting magnitude. With the reference image subtracted, the very wide field camera images do not show the optical counterpart (see Table 1 and Figure 2b) at the NGC 4993 position, down to the V-band limiting magnitude of 15.5 mag.

As the LIGO/Virgo G298048 BAYESTAR-HLV (LIGO Scientific Collaboration and Virgo Collaboration 2017abc) error region was rapidly setting, MASTER-OAFA only had a ~ 3.5h window to observe the error field. Starting from 2017-08-17 22:58:48 UT, MASTER-OAFA observed the new BAYESTAR-HLV localization map of LIGO/Virgo G298048 with unfiltered images (180 sec exposures), down to a limiting magnitude of 19-20$^m$. The co-added images (n frames added) have a fainter limiting magnitude of 20.5.

The final BAYESTAR-HLV localization map is highly elongated, and therefore fields with different priorities set at different times. A special program *"MASTER-Net scheduler"* distributed the sequence of survey images in such a way as to maximize the probable total observing time of the area inside the localization region before it set. In this particular case the MASTER-OAFA telescope was the only MASTER telescope observing the localization region, but the *"MASTER-Net scheduler"* is designed to ensure the fastest possible coverage of the GW error area by the MASTER telescope network. Figures 2c) and 2d) show the dynamics of the inspection survey and the coverage map on the first night since the beginning of observation until the last possible field inside BAYESTAR -HLV localization map disappears beyond the horizon.

During the survey of the GW error region the MASTER-OAFA robotic telescope took two images of the galaxy NGC 4993 and our auto-detection system discovered MASTER OTJ130948.10-232253.3 , also discovered by Swope telescope and called Sss17a (later also named AT2017gfo) and first published in Coulter et al. 2017a and confirmed by multi-wavelength observations, see Abbott et al. 2017b). MASTER images were taken approximately 1 hour after the start of the survey, at 2017-08-17 23:59:54 UT i.e. 11 h 18m 50 sec after the LVC trigger time (Lipunov et al. 2017c,e).

As is evident from Figure 2e, the OT is visible in the subtracted image. The MASTER network software is constantly being improved and for some types of objects, for which it is possible to define several unquestionable automatic verification criteria, the discoveries are published automatically. Examples include GRBs registered by the Swift satellite or NEO asteroid hazards. For example, the optical counterpart for GRB 161017A was observed, detected and a discovery notice automatically published on the GCN approximately 200 s after the notice time (see Yurkov et al. 2016).

However, other types of objects require manual checks before publishing. G298048 was observed when it was the middle of the night in Moscow, and we detected this OT by inspecting software reports only in the morning of 18 August, i.e. several hours after another team had reported it (Coulter et al 2017, Abbott 2017b).

Once the optical source was found, the MACTER-net telescopes stopped the inspection survey inside the LIGO/Virgo probability map, and focused on observing MASTER OT J130948.10-232253.3/ SSS17a (Figure 2b,d,e).

For us the most important argument that this OT in NGC4993 was connected with GW170817, was its unusual spectral properties (Chen et al. 2017, Shara et al 2017, Abbott et al.2017b). The MASTER network has detected about ~1500 optical transients (OTs) during last several years covering 10 different classes of astrophysical types: GRB optical counterparts (MASTER is the leader of the world prompt optical GRB observations), supernovae, novae, dwarf novae, other cataclysmic variable, AGN and QSO flares, as well as anti-transients (dramatically decreased brightness of the stars), comets, Potentially Hazard Asteroids (PHA), ~15 minutes durations UVCet flares and very short OTs of unknown nature (in two tubes simultaniuosly). The OT in NGC4993 was really unlike any of our transients discovered before. We decided that the appearance of this really unusual object, in the LIGO/VIRGO error box 0.5 days after GW event, was not be random , and we have focused on the photometry of this unusual object over the following nights.

Over the next three days both southern MASTER robotic telescopes, MASTER-SAAO and MASTER-OAFA, monitored the possible kilonova, which remained visible for the two instruments, in the B- and R-band filters and in unfiltered mode. Table 1 presents the results of photometry, which was made on subtracted images free from the galaxy background.

As of 1 September 2017, MASTER-OAFA and MASTER-SAAO made a total of more than 600 exposures of the MASTER OT J130948.10-232253.3/Sss17a region.
A video of the coverage map for the first day is presented in Supplement 1.
MASTER OT J130948.10-232253.3/Sss17a photometry is listed in Table 1and Figure 3 and is described in the following Section.

## V. PHOTOMETRY.

The usual MASTER auto-detection software identifies objects (new or known ones) automatically and undertakes photometry using calibrations provided by thousands of USNO-B1 stars in our wide field of view (Lipunov et al 2010), in real time (1-2 minutes after CCD readout), and can work in alert, inspect and survey mode, independently of human intervention. This is our unique feature which gives us the possibility to detect new objects in large fields in real time and to study outbursts in the early stages of explosion (Troja et al. 2017, Lipunov et al. 2016). Later, for accurate photometry, we undertake the following procedure.

It is impossible to produce photometry of this object using standard aperture or PSF photometry methods on the original image due to the galaxy background. For accurate photometry we undertake the following procedure. For the subtraction procedure, we choose the most suitable reference image (by the average star FWHMs and the frame detection limit) from our archive. After a very accurate (sub-pixel) centering of the source and reference images, we obtain a difference image following the technique described in Alard (2000). Using the original image, we determine the transformation of the instrumental flux into standard stellar magnitudes and then we measure the object's instrumental flux from the difference image. We correct the obtained stellar magnitudes for the Galactic extinction, based on E (B-V) = 0.1 mag (Schlafly & Finkbeiner 2011).

The MASTER-NET archive contains 126 images of the galaxy NGC 4993, obtained from 2015-01-17 00:45:46 to 2017-05-02 22:17:04, none of which shows any optical activity for SSS17a, see images' table at Appendix.

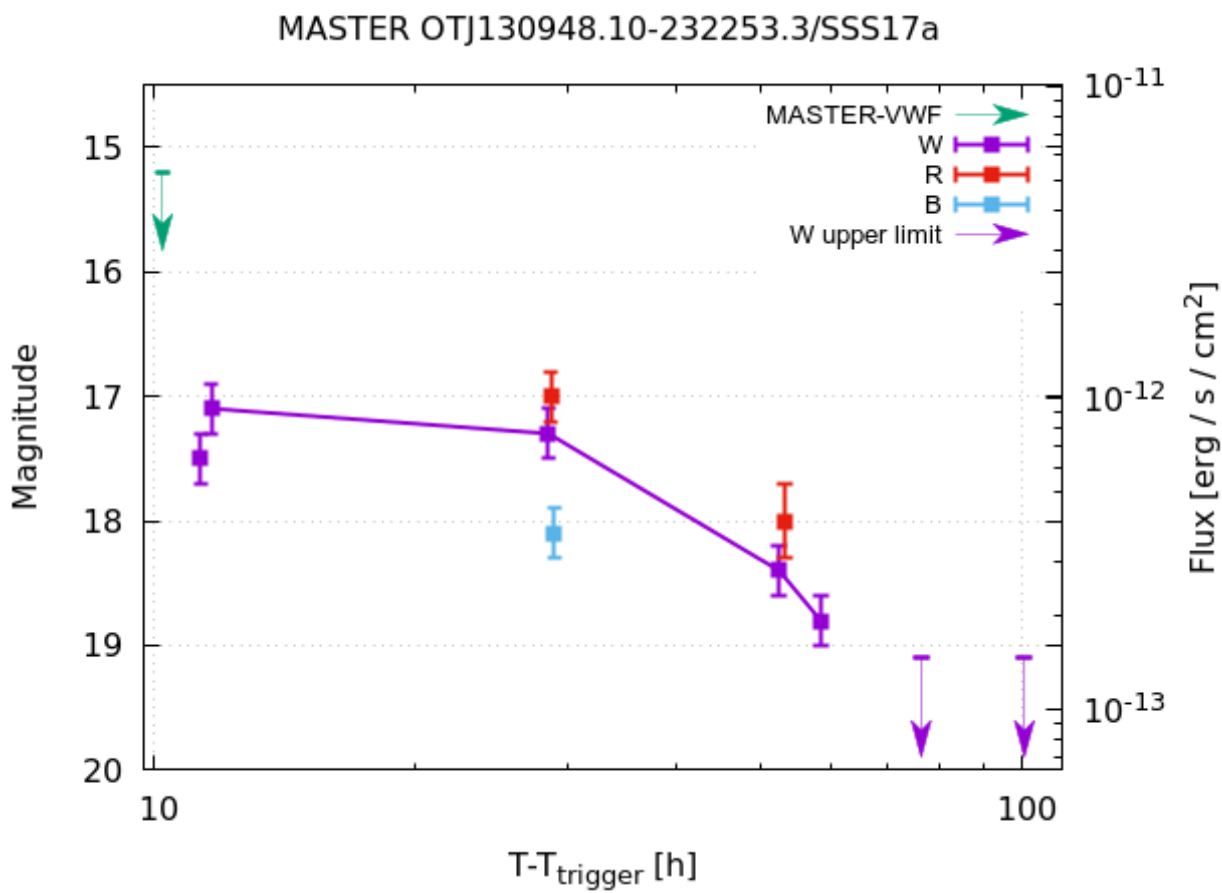

**Figure 3.** MASTER-Net light curve of Kilonova GW170817 in NGC 4993 (see Table 1).

**Table 1.** MASTER photometry of MASTER OTJ130948.10-232253.3/Sss17a ( Galaxy extinction included). Photometry was made on MASTER very wide field cameras (VWFC) and on main MASTER telescope(as you also see from the limit value)

| Date | UT start | Tstart - Ttrig , Kilosec | Tmid - Ttrig , Kilosec | Exp, sec | Filter | Mag. | Err. Mag. | Upper limit | MASTER Site |
|---|---|---|---|---|---|---|---|---|---|
| 2017-08-17 | 22:54:18 | 36.794 | 37.356 | 225x5 | V | >15.2 | --- | 15.5 | MASTER-OAFA-VWFC |
| 2017-08-17 | 23:59:54 | 40.730 | 40.820 | 180 | W | 17.5 | 0.2 | 19.5 | MASTER-OAFA |
| 2017-08-18 | 00:19:05 | 41.881 | 41.971 | 180 | W | 17.1 | 0.2 | 19.3 | MASTER-OAFA |
| 2017-08-18 | 17:06:55 | 102.352 | 102.653 | 6x180 | W | 17.3 | 0.2 | 20.0 | MASTER-SAAO |
| 2017-08-18 | 17:17:33 | 102.989 | 103.463 | 3x180 | R | 17.0 | 0.2 | 19.8 | MASTER-SAAO |
| 2017-08-18 | 17:34:02 | 103.979 | 104.290 | 3x180 | B | 18.1 | 0.1 | 19.5 | MASTER-SAAO |
| 2017-08-19 | 17:06:57 | 188.753 | 189.047 | 3x180 | W | 18.4 | 0.2 | 20.0 | MASTER-SAAO |
| 2017-08-19 | 17:53:34 | 191.550 | 191.844 | 3x180 | R | 18.0 | 0.3 | 19.8 | MASTER-SAAO |
| 2017-08-19 | 18:04:32 | 192.208 | 192.503 | 3x180 | B |  | --- | 19.5 | MASTER-SAAO |
| 2017-08-19 | 23:13:20 | 210.736 | 211.785 | 10x180 | W | 18.8 | 0.2 | 20.7 | MASTER-OAFA |
| 2017-08-20 | 17:04:36 | 275.012 | 275.306 | 3x180 | W | >19.1 | --- | 20.0 | MASTER-SAAO |
| 2017-08-20 | 17:25:56 | 276.292 | 276.586 | 3x180 | R | >18.6 | --- | 19.5 | MASTER-SAAO |
| 2017-08-20 | 17:36:32 | 276.928 | 277.222 | 3x180 | B | >19.3 | --- | 20.0 | MASTER-SAAO |
| 2017-08-21 | 00:26:31 | 301.527 | 302.577 | 10x180 | W | >19.8 | --- | 20.7 | MASTER-OAFA |
| 2017-08-21 | 17:08:14 | 361.630 | 361.925 | 3x180 | W | >19.1 | --- | 20.0 | MASTER-SAAO |
| 2017-08-21 | 18:06:12 | 365.108 | 365.403 | 3x180 | R | >18.6 | --- | 19.5 | MASTER-SAAO |
| 2017-08-21 | 19:20:23 | 369.559 | 369.854 | 3x180 | B | >18.3 | --- | 19.0 | MASTER-SAAO |

* This is the photometry of the faint stellar image derived from image subtraction, removing the galaxy contribution.

Ttrig - LVC trigger Time
Tstart – MASTER exposure start time
Tmid -MASTER exposure middle time
W = 0.8R + 0.2B – unfiltered with respect to USNO-B1 stars.
VWFC - Very Wide Field Camera, MASTER-SAAO and MASTER-OAFA – main MASTER telescopes at these observatories.

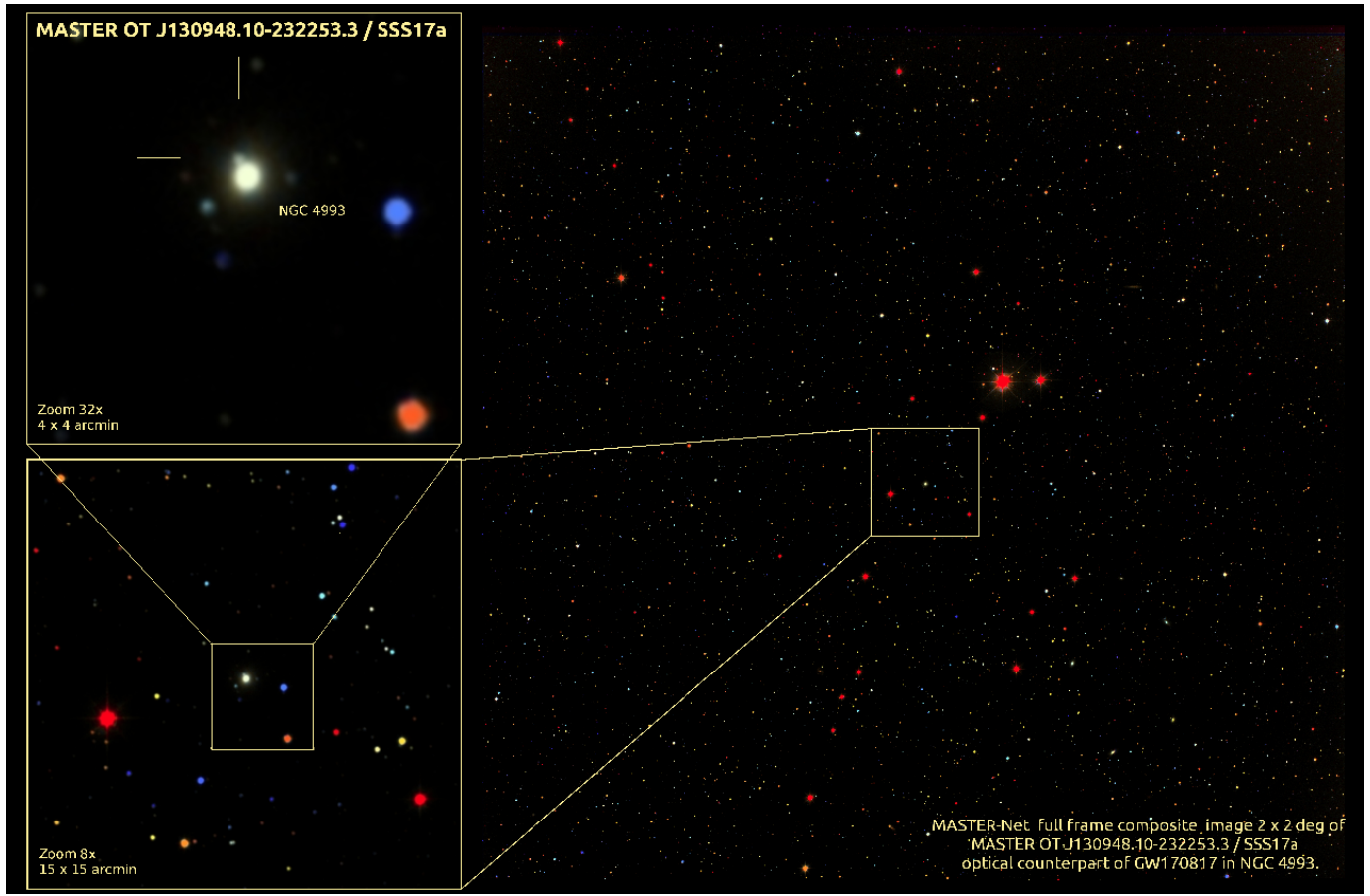

**Figure4** MASTER composed discovery image started 17 Aug 2017 at 23:59:54 UT.We used color B, R, I, W filters MASTER-OAFA and MASTER-SAAO images. The kilonova position is marked by white lines on the left part of composed image. The right (big) image is usual MASTER main telescope's field of view.

In Figure 4 we present the MASTER composed discovery image with kilonova position.

## VI. DISCUSSION

The detection of electromagnetic radiation accompanying the coalescence of neutron stars was by no means a surprise. The merger of neutron stars as a formation mechanism of gamma-ray bursts was first considered by Blinnikov et al. in 1984, and the occurrence rate of such events was computed back in 1987 using the population synthesis method (Scenario Machine) by Lipunov et al. (1987) and later refined by taking into account the evolution of star formation for the entire Universe (Lipunov et al., 1995).

It was found that neutron-star mergers in a Milky-Way type galaxy (i.e. in a galaxy with the mass of $10^{11}$ solar masses and star formation of one solar mass per year) occur at a rate of $\sim 10^{-4}$ yr$^{-1}$ (Lipunov et al., 1987; see the caption to Figure 2, case "e", in that paper). It immediately follows from this that in our neighborhood one merger per year should occurs within the sphere containing 10,000 galaxies of the Milky Way type. This volume corresponds to the radius of D ~ 20 Mpc . Hence the first Monte Carlo population synthesis in 1987 predicted an occurrence rate of several events per year within the sphere of radius ~ 40 Mpc, which agreed with 2017 observations quite well. Note that the work of Phinney (1991) using estimation from pulsar binary period deviation, gave two orders of lower rate for merging NS.

The proximity of GW170817 is entirely consistent with the evolution of NS+NS merger rates in the Universe later computed using the Scenario Machine (Lipunov et al., 1995,; Lipunov &Pruzhinskaya 2014; cf. Belczynski et al., 2002).

Even if the line of sight to gamma-ray burst is not co-aligned with the jet during the merger (the corresponding probability is close to 99.9 %), the event may still produce more isotropic accompanying, or even leading, radiation.

The possibility of a kilonova explosion resulting from the decay of radioactive elements in the expanding envelope ( Li, L.-X. & Paczynski, B., 1998; Rosswog, S. et al.,1999; Freiburghaus et al.,, 1999; Rosswog, S. et al., 2005; Metzger, B. D. et al, 2010; Coughlin et al., 2017).

A number of very important arguments supporting the hypothesis that SSS17a is a kilonova candidate were published from analysis of the wideband photometry (see Abbot et al., 2017b) and spectral observations (Drout, et al., 2017; Shara et al., 2017, Abbot et al., 2017b) .

Let us now compare the observed optical luminosity with the optical luminosities of other possible kilonovae observed earlier, namely, for the gamma-ray bursts GRB 130603B (Tanvir et al., 2013) and GRB 080503 (Perley et al., 2009), see Table 2. For the maximum brightness of the kilonova in NGC 4993 we used first point on the MASTER light curve (Figure 3). We use the redshifts for the two galaxies that are visually closest to the GRB position (see discussion below).

**Table 2.** Possible kilonova brightness. There are name

| Name | Z | DL Mpc | DM | Obs | Band | $m_{vis}$ Max[AB] | Mabs Flat spec AB | Flux iso erg/s | link |
|---|---|---|---|---|---|---|---|---|---|
| **Kilonova NGC 4993** | 0.0098 | 42.5 | 33.14 | MASTER | W | 17.3 | -16.03 | **$10^{42}$** | This paper |
| **GRB 130603** | 0.3560 | 1911.9 | 41.41 | HST | H | 25.73 | -15.35 | **$5.50 \times 10^{41}$** | Tanvir et al (2013) |
| **GRB 080503** | 0.561* | 3290.5 | 42.59 | Gemini/ Keck | r | 25.48 | -16.62 | **$1.78 \times 10^{42}$** | Perley et al (2009) |

The source spectra are unknown and following Perley et al. (2009, Figure 7) we use a flat spectrum to estimate the k correction. Hence M=m−DM+2.5 log(1 +z), where DM is the distance modulus. Furthermore, for flat spectrum sources the absolute magnitudes in all bins are equal.

We assume that the redshift of GRB 080503 is equal to the redshifts of the two galaxies visually closest to it. However, there still is not a consensus regarding the host galaxy of GRB 080503. Perley et al (2009) suggest that the nearest galaxies have no connection with GRB 080503, due to large angular separations. However, we believe that any of these galaxies could potentially be a host because NS+NS systems must obtain a huge kick velocity during the two SN explosions, allowing them to escape from the galaxy before the collision. In any case, this is the unique and best estimate of the redshift for a given object.

The agreement between the observed absolute magnitudes and characteristic luminosities evoke the old idea about viewing gamma-ray bursts as standard candles (Lipunov et al. 2001). However, we are now dealing with kilonovae that accompany short gamma-ray burst events. Here we have rather an analogy with Type Ia supernovae. Both kinds of event may represent collisions of compact stars: binary white dwarfs and binary neutron stars in the case of supernovae and kilonovae, respectively. In addition, the mass of the collapsing object may also play an important part in both processes. The Chandrasekhar limit for SN Ia and the Oppenheimer—Volkov limit for kilonovae. Of course this reasoning may be too naïve given that the densities of objects differ by a million fold. Also the kilonova is fainter than SNIa and do not have simple spectral lines. Due to of the kick velocity, double neutron stars systems can escape from their host galaxies. In this case, the brightness of the kilonovae is the only one indicator of its distance.

Furthermore, optical radiation plays an important part in the energy balance of SN Ia, whereas this is by no means evident in the case of kilonovae, for which the total energy release in the IR and X-ray parts of the spectrum can be more important. In this connection, of interest is the discovery of X-ray emission from kilonova in NGC4993 (Troja E. et al.,2017), which provides new information about the astrophysical properties of the kilonovae.

**Acknowledgments**

MASTER project is supported in parts by the Development Programm of Lomonosov Moscow State University, Moscow Union OPTICA, Russian Science Foundation 16-12-00085; and National Research Foundation of South Africa. NB was supported in parts by RFBR 17-52-80133, Russian Federation Ministry of Education and Science (14.B25.31.0010, 14.593.21.0005); AG by RFBR 15-02-07875.
We are especially grateful to S.M.Bodrov for his long years MASTER's support and to Dmitry Svinkin for collaboration.

**Appendix**

MASTER-NET 126 prediscovery images of G298048 optical counterpart since 2015-01-17 00:45:46 till 2017-05-02 22:17:04UT

| Datetime, UT | Exp.time, s | Filter | Upper_limit |
|---|---|---|---|
| 2017-05-02 22:17:04.118 | 60 | W | 19.73 |
| 2017-05-02 22:03:32.383 | 60 | W | 19.70 |
| 2017-04-19 00:05:48.714 | 60 | W | 19.43 |
| 2017-04-18 23:54:57.474 | 60 | W | 19.49 |
| 2017-03-10 22:39:05.35 | 60 | W | 18.79 |
| 2017-03-10 22:39:05.349 | 60 | W | 18.70 |
| 2017-03-10 22:26:22.364 | 60 | W | 18.80 |
| 2017-03-10 22:26:22.343 | 60 | W | 18.70 |
| 2017-03-10 22:15:13.272 | 60 | W | 18.73 |
| 2017-03-10 22:15:13.25 | 60 | W | 18.70 |
| 2017-02-08 01:25:35.584 | 60 | W | 19.50 |
| 2017-02-08 01:14:42.231 | 60 | W | 19.50 |
| 2016-12-16 02:36:26.611 | 60 | W | 16.92 |
| 2016-12-16 02:36:26.59 | 60 | W | 17.00 |
| 2016-12-16 02:04:04.222 | 60 | W | 17.94 |
| 2016-12-16 02:04:04.21 | 60 | W | 17.90 |
| 2016-08-28 17:34:48.436 | 60 | W | 18.93 |
| 2016-08-28 17:33:09.212 | 60 | W | 19.04 |
| 2016-08-28 17:27:45.208 | 60 | W | 19.10 |
| 2016-08-28 17:27:45.208 | 180 | W | 19.64 |
| 2016-08-28 17:26:01.303 | 60 | W | 19.06 |
| 2016-08-28 17:20:34.987 | 60 | W | 18.65 |
| 2016-08-28 17:18:57.252 | 180 | W | 19.44 |
| 2016-08-28 17:18:57.252 | 60 | W | 18.60 |
| 2016-08-20 18:01:44.034 | 60 | W | 19.40 |
| 2016-08-20 17:56:18.846 | 60 | W | 19.42 |
| 2016-08-20 17:50:54.677 | 60 | W | 19.32 |
| 2016-06-21 18:53:44.099 | 60 | W | 19.21 |
| 2016-06-21 18:53:44.096 | 60 | W | 19.10 |
| 2016-06-21 18:43:23.665 | 60 | W | 19.14 |
| 2016-06-21 18:43:23.626 | 60 | W | 19.21 |
| 2016-06-21 18:32:58.607 | 60 | W | 19.21 |
| 2016-06-21 18:32:58.605 | 60 | W | 19.10 |
| 2016-06-17 21:53:12.354 | 60 | W | 18.56 |
| 2016-06-17 21:42:16.583 | 60 | W | 18.51 |
| 2016-06-17 21:31:33.343 | 60 | W | 18.50 |
| 2016-05-11 19:16:34.03 | 60 | W | 19.74 |
| 2016-05-11 19:16:34.003 | 60 | W | 19.67 |
| 2016-05-11 19:03:15.997 | 60 | W | 19.77 |
| 2016-05-11 19:03:15.994 | 60 | W | 19.65 |
| 2016-05-11 18:51:12.122 | 60 | W | 19.73 |
| 2016-05-11 18:51:12.097 | 60 | W | 19.70 |
| 2016-05-05 20:20:07.211 | 180 | W | 20.00 |
| 2016-05-05 20:16:37.894 | 180 | W | 20.05 |
| 2016-05-05 20:13:09.175 | 180 | W | 20.00 |
| 2016-05-05 19:47:44.764 | 180 | W | 20.16 |
| 2016-05-05 19:44:10.718 | 180 | W | 20.13 |
| 2016-05-05 19:40:37.21 | 180 | W | 20.16 |
| 2016-05-05 19:14:58.806 | 180 | W | 20.53 |
| 2016-04-12 01:36:46.999 | 180 | W | 19.75 |
| 2016-04-12 01:36:46.996 | 180 | W | 19.58 |
| 2016-04-12 01:32:18.804 | 180 | W | 19.86 |
| 2016-04-12 01:32:18.799 | 180 | W | 19.60 |
| 2016-04-12 01:17:06.783 | 180 | W | 19.70 |
| 2016-04-12 01:17:06.781 | 180 | W | 19.67 |
| 2016-04-12 01:17:06.781 | 540 | W | 19.92 |
| 2016-04-12 00:59:15.746 | 180 | W | 20.06 |
| 2016-04-12 00:59:15.743 | 180 | W | 20.10 |
| 2016-02-21 00:29:56.411 | 60 | W | 19.21 |
| 2016-02-21 00:19:43.054 | 60 | W | 19.36 |
| 2016-02-21 00:09:02.759 | 60 | W | 19.39 |
| 2015-12-15 01:37:02.481 | 60 | W | 19.40 |
| 2015-12-15 01:37:02.477 | 60 | W | 19.15 |
| 2015-12-15 01:30:20.686 | 60 | W | 19.36 |
| 2015-12-15 01:30:20.684 | 60 | W | 19.09 |
| 2015-12-15 01:25:16.74 | 60 | W | 19.30 |
| 2015-12-15 01:25:16.729 | 60 | W | 19.07 |
| 2015-11-29 02:32:56.228 | 60 | W | 15.74 |
| 2015-11-29 02:32:56.197 | 60 | W | 15.60 |
| 2015-11-29 02:29:35.963 | 60 | W | 16.27 |
| 2015-11-29 02:29:35.937 | 60 | W | 15.91 |
| 2015-11-29 02:28:00.948 | 60 | W | 16.03 |
| 2015-11-29 02:28:00.93 | 60 | W | 16.42 |
| 2015-11-29 02:26:25.431 | 60 | W | 16.55 |
| 2015-11-29 02:26:25.431 | 180 | W | 17.17 |
| 2015-11-29 02:26:25.428 | 180 | W | 16.74 |
| 2015-11-29 02:26:25.428 | 60 | W | 16.16 |
| 2015-08-29 17:44:19.853 | 60 | W | 17.39 |
| 2015-08-29 17:37:59.363 | 60 | W | 17.45 |
| 2015-08-29 17:31:31.398 | 60 | W | 17.65 |
| 2015-08-29 17:29:59.822 | 60 | W | 17.76 |
| 2015-08-29 17:29:59.822 | 180 | W | 18.39 |
| 2015-06-30 20:25:08.182 | 60 | W | 18.75 |
| 2015-06-30 20:25:08.178 | 60 | W | 18.48 |
| 2015-06-30 20:13:55.063 | 60 | W | 18.84 |
| 2015-06-30 20:13:55.02 | 60 | W | 18.61 |
| 2015-06-30 20:02:44.457 | 60 | W | 17.80 |
| 2015-06-30 20:02:44.434 | 60 | W | 18.05 |
| 2015-06-20 18:15:48.359 | 60 | W | 19.94 |
| 2015-06-20 18:04:13.291 | 60 | W | 19.91 |
| 2015-06-20 17:52:43.813 | 60 | W | 19.80 |
| 2015-04-24 20:29:52.994 | 60 | W | 19.73 |
| 2015-04-24 20:29:52.966 | 60 | W | 19.56 |
| 2015-04-24 20:23:26.48 | 60 | W | 19.56 |
| 2015-04-24 20:23:26.476 | 60 | W | 19.73 |
| 2015-04-24 20:17:15.845 | 60 | W | 19.53 |
| 2015-04-24 20:17:15.791 | 60 | W | 19.77 |
| 2015-03-25 20:49:45.774 | 60 | W | 19.69 |
| 2015-03-25 20:49:45.757 | 60 | W | 19.77 |
| 2015-03-25 20:36:26.963 | 60 | W | 19.72 |
| 2015-03-25 20:36:26.946 | 60 | W | 19.60 |
| 2015-03-25 20:25:13.801 | 60 | W | 19.61 |
| 2015-03-25 20:25:13.765 | 60 | W | 19.50 |
| 2015-02-26 22:27:08.995 | 60 | W | 19.44 |
| 2015-02-26 22:27:08.323 | 60 | W | 19.47 |
| 2015-02-26 22:12:35.418 | 60 | W | 19.43 |
| 2015-02-26 22:12:35.357 | 60 | W | 19.35 |
| 2015-02-26 21:51:36.812 | 60 | W | 19.43 |
| 2015-02-26 21:51:36.609 | 60 | W | 19.32 |
| 2015-01-21 00:17:56.9 | 60 | W | 19.75 |
| 2015-01-20 00:18:10.584 | 60 | W | 19.73 |
| 2015-01-20 00:18:10.536 | 60 | W | 19.70 |
| 2015-01-20 00:16:30.601 | 60 | W | 19.63 |
| 2015-01-20 00:16:30.455 | 60 | W | 19.64 |
| 2015-01-20 00:14:52.539 | 60 | W | 19.68 |
| 2015-01-20 00:14:52.539 | 180 | W | 20.22 |
| 2015-01-20 00:14:52.439 | 60 | W | 19.60 |
| 2015-01-20 00:14:52.439 | 180 | W | 20.16 |
| 2015-01-17 00:49:44.307 | 60 | W | 19.43 |
| 2015-01-17 00:49:44.121 | 60 | W | 19.30 |
| 2015-01-17 00:47:41.118 | 60 | W | 19.10 |
| 2015-01-17 00:47:40.16 | 60 | W | 18.95 |
| 2015-01-17 00:45:46.426 | 180 | W | 19.90 |
| 2015-01-17 00:45:46.426 | 60 | W | 19.50 |
| 2015-01-17 00:45:46.291 | 60 | W | 19.36 |
| 2015-01-17 00:45:46.291 | 180 | W | 19.73 |

## REFERENCES

Abbott, B., et al., 2016a. ApJ 818L, 22A
Abbott, B., et al., 2016b. Phys. Rev. Lett. 116, 1102.
Abbott, B., et al., 2016c. ApJL 826, 13A
Abbott, B. et al. 2017a Phys Rev, in press
Abbott, B. et al. 2017b, ApJL-16-Oct-2017
Alard, C. 2000 A&AS, 144, 363
Allam S. et al., 2017, LVC GCN 21530, 1
Bartos et al., 2017, GCN, 1
Berger, E.; Fong, W.; Chornock, R. 2013, ApJ, 774L, 23B
Belczynski, K., Kalogera, V., & Bulik, T. 2002, ApJ, 572, 407
Blinnikov, S.I., Novikov, I.D., Perevodchikova, T.V. & Polnarev, A.G. 1984, SvAL, 10, 177
Chambers K.C., Huber M.E., Smartt S.J et al. 2017, GCN, 21553
T.-W. Chen, P. Wiseman, J. Greiner, and P. Schady, D'Elia et al., 2017, GCN, 21592, 1
Clark, J. P. A.; van den Heuvel, E. P. J.; Sutantyo, W. 1979, A&A, 72, 120C
Cook D., Van Sistine A., Singer L., Kasliwal M. M. 2017a, GCN 21519, 1
Coughlin M., et al., 2017 arXiv:1708.07714v1
Coulter, C. D. Kilpatrick, M. R. Siebert et al. 2017, GCN, 21529, 1
Drout, J. D. Simon, B. J. Shappee et al., 2017, GCN, 21547,1
Freiburghaus, C., Rosswog, S. & Thielemann, F.-K. 1999, ApJ, 525, L121
Gorbvskoy E., Ivanov K., Lipunov V., et al. 2010, AdAst, 2010, 62G
Gorbovskoy E., Lipunov V., Kornilov V., et al. 2013 ARep, 57, 233
Kasliwal et al. 2017, DOI 10.1126/science.aap9455;
von Kienlin A., Meegan C., Goldstein A. 2017, GCN, 21520
Kornilov V., Lipunov V., Gorbovskoy E., et al. 2012, ExA, 33, 173
Li, L.-X. & Paczynski, B. Transient events from neutron star mergers. 1998, ApJ, 507, L59
LIGO Scientific Collaboration and Virgo Collaboration (Essik et al.) 2017a, GCN, 21505, 1
LIGO Scientific Collaboration and Virgo Collaboration (Connaughton et al.) 2017b, GCN, 21506, 1
LIGO Scientific Collaboration and Virgo Collaboration (Singer et al.) 2017c, GCN, 21513, 1
Lipunov V.M., Gorosabel J., Pruzhinskaya M., et al. 2016, MNRAS, 455, 712
Lipunov, V. M.; Kornilov, V.; Gorbovskoy, E. et al. 2017a MNRAS, 465, 3656L
Lipunov, V. M.; Kornilov, V.; Gorbovskoy, E. et al. . 2017b, NewA, 51, 122
Lipunov, V., Gorbovskoy, E., Kornilov V., et al., 2017c, GCN, 21546, 1
Lipunov, V., Gorbovskoy, E., Kornilov V., et al., 2017d, GCN, 21570, 1
Lipunov, V., Gorbovskoy, E., Kornilov V., et al., 2017e, GCN, 21587, 1
Lipunov, V., Gorbovskoy, E., Kornilov V., et al., 2017f, GCN, 21687, 1
Lipunov V., 2017g, GCN, 21621, 1
Lipunov V., Kornilov V., Gorbovkoy E., et al. 2010, AdAst, 2010, 30L
Lipunov et al., 1995, ApJ 454, 593
Lipunov V & Pruzhinskaya M, 2014, MNRAS, 440, 1193
Lipunov, V. M.; Panchenko, I. E. 1996 A&A, 312, 937L
Lipunov, V. M.; Gorbovskoy, E. S., 2008, MNRAS, 383, 1397L
Lipunov, V.M., Postnov, K.A. & Prokhorov, M.E., 2001, ARep, 45, 236
Lipunov, V. M.; Postnov, K. A.; Prokhorov, M. E.1987, A&A, 176, L1
Lipunova, G. V.; Gorbovskoy, E. S.; Bogomazov, A. I.; Lipunov, V. M. 2009, MNRAS, 397, 1695L
Lipunova, G. V.; Lipunov, V. M., 1998, A&A, 329L, 29L
Lyman J, D. Homan (Univ. of Edinburgh), K. Maguire et al GCN, 21582, 1
Metzger, B. D.; Martínez-Pinedo, G.; Darbha, S. 2010, MNRAS 406, 2650
Pian E., V. D'Elia, S. Piranomonte et al GCN, 21592, 1
Perley D.A. et al., 2009, ApJ, 696, 1871
Phinney, E.S., 1991. Astrophys. J. Lett. 380, L17–L21.
Rosswog, S. et al. 1999, A&A 341, 499
Rosswog, S2005 AJ 634, 1202
Savchenko V., Mereghetti S., Ferrigno C. 2017, GCN, 21513, 1
Schlafly & Finkbeiner 2011, ApJ, 737 (2), article id. 103
Shara, T. Williams, P. Vaisanen et al, GCN, 21610, 1
Singer, Leo P.; Price, Larry R. 2016 PhysRev D, 93 (2), id.024013
Singer, L.P., Chen, H.-Yu., Holz, D.E. et al. 2016 AJL, 829(1), L15
Tanvir et al., 2013, Nature, 500, 547
Troja, E., Lipunov, V., Mundel, C. et al., 2017a Nature, 547, 425
Troja, E., Piro, L., Sakamoto, T. et al 2017b, GCN, 21765
Yurkov, V., Sergienko, Yu., Varda, D. et al. 2016 GCN 20063, 1

SUPPLEMENT 1.

http://observ.pereplet.ru/cams/natasha/MASTER.G298048.180.24b.mp4

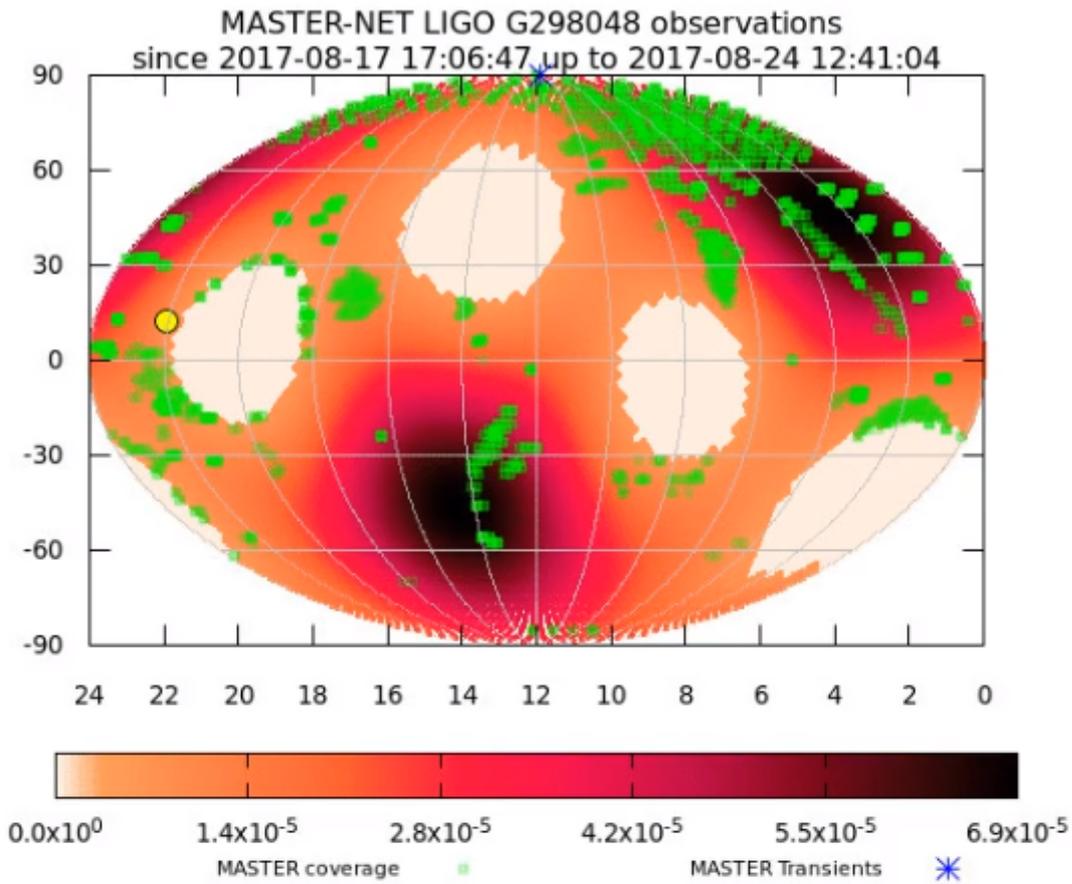

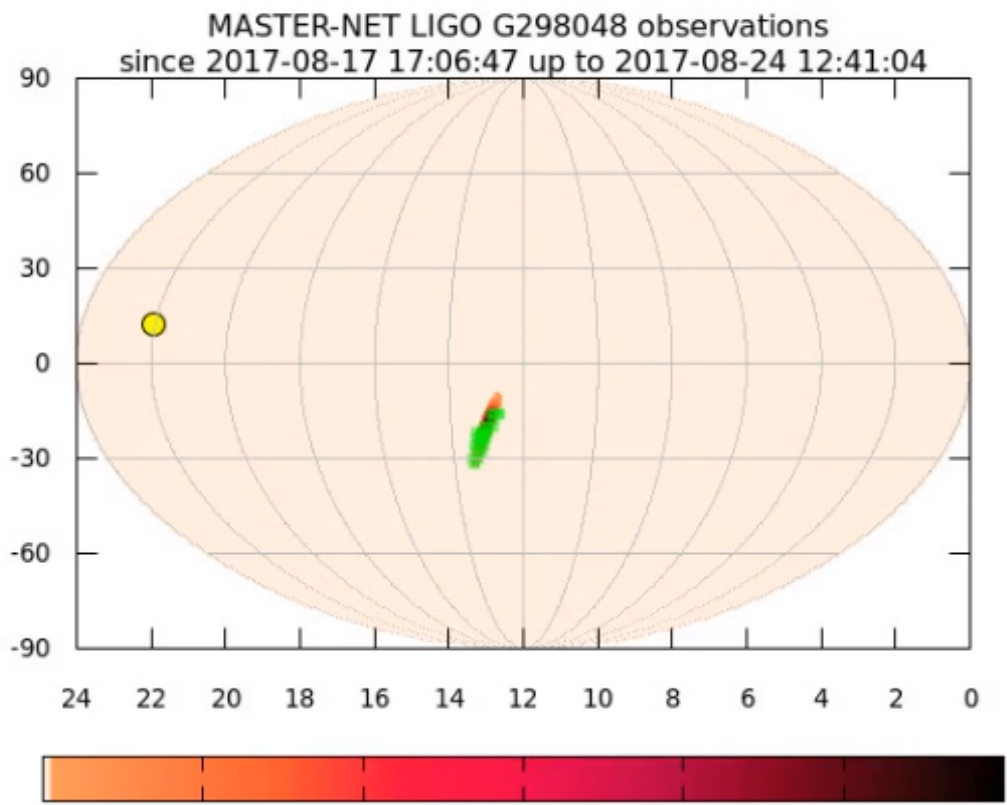